\documentclass[aps,prb,showpacs,groupedaddress,twocolumn]{revtex4}
\usepackage{amsmath}
\usepackage{amssymb}
\usepackage{graphicx}
\usepackage{txfonts}
\usepackage{color}
\usepackage{graphics}
\usepackage{ams}
\usepackage{ulem}



\begin{document}
\title{Nernst and Seebeck effect in a graphene nanoribbon}

\author{Yanxia Xing$^1$, Qing-feng Sun$^{2}$,
and Jian Wang$^{1}$}

\address{$^1$Department of Physics and the Center of Theoretical and
Computational Physics, The University of Hong Kong, Pokfulam Road,
Hong Kong, China
\\
$^2$Beijing National Lab for Condensed Matter Physics and Institute
of Physics, Chinese Academy of Sciences, Beijing 100080, China}

\begin{abstract}
The thermoelectric power, including the Nernst and Seebeck
effects, in graphene nanoribbon is studied. By using the
non-equilibrium Green function combining with the tight-binding
Hamiltonian, the Nernst and Seebeck coefficients are obtained. Due
to the electron-hole symmetry, the Nernst coefficient is an even
function of the Fermi energy while the Seebeck coefficient is an
odd function regardless of the magnetic field. In the presence of
a strong magnetic field, the Nernst and Seebeck coefficients are
almost independent of the chirality and width of the nanoribbon,
and they show peaks when the Fermi energy crosses the Landau
levels. The height of $n$-th (excluding $n=0$) peak is
$[\ln2/|n|]$ for the Nernst effect and is $\ln2/n$ for the Seebeck
effect. For the zeroth peak, it is abnormal with height $[2\ln2]$
for the Nernst effect and the peak disappears for the Seebeck
effect. When the magnetic field is turned off, however, the Nernst
effect is absent and only Seebeck effect exists. In this case, the
Seebeck coefficient strongly depends on the chirality of the
nanoribbon. The peaks are equidistant for the nanoribbons with
zigzag edge but are irregularly distributed for the armchair edge.
In particular, for the insulating armchair ribbon, the Seebeck
coefficient can be very large near the Dirac point. When the
magnetic field varies from zero to large values, the differences
among the Seebeck coefficients for different chiral ribbons
gradually vanish and the nonzero value of Nernst coefficient
appears first near the Dirac point then gradually extents to the
whole energy region.
\end{abstract}
 \maketitle

\section{introduction}
As a single atomic layer extracted from graphite, graphene has been
successfully fabricated experimentally.\cite{ref1,ref01} Due to its
peculiar topological structure, the graphene exhibits peculiar
properties.\cite{ref2} For the graphene sheet, the conduction and
valence band in graphene intersect at Dirac points, the corners of
the hexagonal first Brillouin zone. Around the Dirac points graphene
has a unique band structure and its quasi particles satisfy the
massless Dirac equation where the speed of light is replaced by the
Fermi velocity of graphene ($v_F\approx10^6m/sec$). Experimentally,
by varying the gate voltage, the charge carriers of graphene can be
easily tuned, globally\cite{ref1} or locally\cite{ref3}. As a result
the Fermi level can be above or below the Dirac points, which is
viewed as electron-like or hole-like system. Along the different
crystal direction in honeycomb lattice, the band
structure\cite{ref4} and the transport properties are different. For
the graphene ribbon with the zigzag edge, a special edge state
exists.\cite{ref5} While for the graphene ribbon with armchair edge,
it is metallic when the transverse layer $N=3M-1$ with integer M and
insulator otherwise.\cite{ref5} When the perpendicular magnetic
field is strong enough to form Landau levels (LLs), these
differences due to different chirality at the zero magnetic field
disappear. In addition, both theoretically\cite{theory2} and
experimentally,\cite{ref2} the Hall conductance was found to be the
half-integer in the values $g (n+1/2) e^2/h$ with degeneracy $g=4$,
indicating that the quantization condition is shifted by a
half-integer compared with the usual integer quantum Hall effect. It
is a direct manifestation of the unique electronic structure of
graphene.

The thermoelectric power (TEP), or the thermal gradient induced
current (or bias with an open boundary), results from a balance of
electric and thermal forces acting on the charge carriers. In
general, we consider two thermoelectric powers, the Nernst effect
which is the transverse TEP induced by a longitudinal thermal
gradient in a perpendicular magnetic field and the Seebeck effect
which is the thermal gradient induced bias in a two probe system.
TEP is of great importance in understanding electronic transport
because it is more sensitive to the details of the density of
states \cite{ref6} and the particle-hole asymmetry\cite{ref7} than
the conductance. In the early days, because of the experimental
difficulty (particularly in low-dimensional systems or
nano-devices), the Nernst effect and Seebeck effect are often
neglected. Instead, one usually measures the Hall effect and the
resistivity. Now, with the development of the micro-fabrication
technology and the low-temperature measurement technology, the
thermoelectric measurement in low-dimensional samples has been
feasible.\cite{ref8} Recently, the Nernst effect and Seebeck
effect have been widely observed and experimentally investigated
in many systems, including the high-Tc
superconductivity,\cite{ref9} ferromagnets,\cite{ref10}
semimetallic,\cite{ref11} Pyrochlore Molybdates,\cite{ref12}
Bismuch,\cite{ref13} single walled carbon nanotube,\cite{ref14}
etc. For the graphene, the study of thermoelectric properties can
elucidate details of the electronic structure of the ambipolar
nature that cannot be realized by probing conductance alone. Very
recently, using a microfabricated heater and thermometer
electrodes, the conductance and the diffusive TEP of graphene are
simultaneously measured by Zuev {\sl et.al.}\cite{Gtherm1} and Wei
{\sl et.al.}.\cite{Gtherm2} Zuev {\sl et.al.} found electrons and
holes contribute to Seebeck effect in opposite ways. At high
temperatures direct measurement of Seebeck coefficient $S_C$ can
be compared with that calculated from the Mott
relation.\cite{Mott} Furthermore, divergence of $S_C$ and the
large Nernst signal were found near the charge neutral point (i.e.
the Dirac point).\cite{Gtherm2} Also, at low temperatures,
% JIAN TEP is oscillating as a function of E?
depending on $E_F$, TEP is oscillating. The temperature suppresses
the oscillation and enhances the magnitude of TEP.

Up to now, some theoretical investigations have been carried out on
the thermal response in the graphene. The electronic transport
coefficients including thermopower was semiclassically treated and
only classical Hall effect (low field) in graphene was
studied.\cite{theory1} In addition, the Nernst coefficient was
studied only in weak magnetic field. It was found to be strong and
positive near Dirac point.\cite{theory2} For a strong magnetic field
in the quantum Hall regime, the Seebeck coefficient was studied and
was focused on its dependence of the field
orientation.\cite{theory3} In all these works, the quantum Nernst
effect is absent because of the calculational subtleties in the
presence of the strong magnetic field. For the normal two
dimensional electron gas characterized by a parabolic dispersion,
the Nernst effect has been
studied.\cite{NernstTheory1,NernstTheory2,KuboNernst} Of these
works, two alternative boundary conditions were considered in
calculating the thermal response functions. One is the adiabatic
boundary condition that the temperatures in the upper and lower edge
are fixed. In this case the Nernst coefficient is similar to the
Seebeck coefficient.\cite{NernstTheory1} The other one is the
non-adiabatic boundary condition on the upper and lower edges, in
which the edge currents are in contact with two heat baths with
different temperatures.\cite{KuboNernst} The Nernst coefficient is
different from the Seebeck coefficient. It is the purpose of our
work to focus on the quantum Nernst effect in the graphene
nanoribbon with the adiabatic boundary condition.

In this paper, we carry out a theoretical study of the Nernst
effect in a crossed graphene nanoribbon and the Seebeck effect in
a single graphene nanoribbon in the strong perpendicular magnetic
field, zero magnetic field, and weak magnetic field. By using the
tight binding model and the nonequilibrium Green function method,
the transmission coefficient and consequently the Nernst and
Seebeck coefficients are obtained. In a strong perpendicular
magnetic field $B$, high degenerated LLs are formed, and the edge
states dominate the transport processes, so the Nernst (Seebeck)
coefficients are almost the same along the different chiral
directions. We find that the Nernst coefficient $N_C$ and the
Seebeck coefficient $S_C$ show peaks when the Fermi energy $E_F$
passes the LLs. At $E_F=0$, because the zeroth LL is shared by
electron-like and hole-like Landau states, $N_C$ which is an even
function of $E_F$ has the highest peak while $S_C$ which is an odd
function of $E_F$ vanishes. On the other hand, at zero $B$, there
is no Lorentz force to bend the trajectories of the thermally
diffusing carriers, so Nernst effect is absent. In this case, the
Seebeck coefficient $S_C$ is strongly dependent on the chirality
of graphene ribbon. In particular, for the insulating armchair
ribbon, $S_C$ can be very large near the Dirac point. At last, the
crossover behavior of the thermoelectric power from the zero
magnetic field to the strong magnetic field is also studied.

The rest of the paper is organized as follows. In Section II, the
models for crossed graphene ribbon or single graphene ribbon are
introduced. The formalisms for calculating the Nernst and Seebeck
coefficient are then derived. Section III gives numerical results
along with discussions. Finally, a brief summary is presented in
Section IV.

\section{model and formalism}

We consider two graphene systems: a four terminal crossed graphene
nanoribbon and a two terminal graphene nanoribbon as shown in the
left and right insets of Fig.1(b). Here we consider ballistic two
dimensional electron gas in which the mean free path and the phase
coherent length are greater than the device size. In the
experiment, we can use the smaller sample to reduce the device
size, and use the lower temperature or the higher magnetic field
to enhance the phase coherent length. In the tight-binding
representation, the Hamiltonian operator can be written in the
following form:\cite{ref2,ref19,ref20}
\begin{eqnarray}
 H_{G} =\sum_{\bf i} \epsilon_{\bf i} a^{\dagger}_{\bf i} a_{\bf i}
      -\sum_{<{\bf ij}>} t e^{i\phi_{ij}} a_{\bf i}^{\dagger} a_{\bf
      j},\label{Ham}
\end{eqnarray}
where ${\bf i}=(i_x , i_y)$ is the index of the discrete honeycomb
lattice site which is arranged as in inset of Fig.1b, and $a_{\bf
i}$ and $a_{\bf i}^{\dagger}$ are the annihilation and creation
operators at the site ${\bf i}$. $\epsilon_{\bf i}$ is the on-site
energy (i.e. the energy of the Dirac point) which can be controlled
experimentally by the gate voltage, here we set $\epsilon_i=0$ as an
energy zero point. The second term in Eq.(2) is the hopping term
with the hopping energy $t$. When the graphene ribbon is under a
uniform perpendicular magnetic field $B_z=B$, a phase $\phi_{ij}$ is
added in the hopping term, and $\phi_{ij}=\int_i^j \vec{A} \cdot
d\vec{l}/\phi_0$ with the vector potential $\vec{A}=(-By,0,0)$ and
the flux quanta $\phi_0=\hbar/e$.

With this ballistic system, the current flowing to the p-th graphene
lead can be calculated from the Landauer-B$\ddot{u}$ttiker
formula:\cite{addnote1}
\begin{eqnarray}
 J_{p}&=&\frac{2e}{\hbar}\sum_q \int\frac{dE}{2\pi}
[ T_{pq}(E)(f_{p}(E)-f_q(E)) ].\label{Landuer}
\end{eqnarray}
where $p,q=1,2,3,4$ for the four terminal system or $p,q=1,2$ for
the two terminal system, and $T_{pq}$ is the transmission
coefficient from terminal-q to terminal-p.

In Eq.(\ref{Landuer}), the transmission coefficient $T_{pq}$ can
be calculated from $T_{pq}(E)=Tr[\Gamma_{p}G^r\Gamma_{q}G^a]$,
where the line-width function
$\Gamma_{p}(E)=i(\Sigma_{p}^r-\Sigma_{p}^{r\dagger})$. The Green's
function
$G^r(E)=[G^a(E)]^{\dagger}=\{EI-H_0-\sum_{p}\Sigma^r_{p}(E)\}^{-1}$
where $H_0$ is Hamiltonian matrix of the central region and $I$ is
the unit matrix with the same dimension as that of $H_0$, and
$\Sigma_{p}^r$ is the retarded self-energy function from the
lead-p. The self-energy function can be obtained from
$\Sigma^r_p(E)=H_{c,p}g^r_p(E)H_{p,c}$, where $H_{c,p}$
($H_{p,c}$) is the coupling from central region (lead-p) to lead-p
(central region) and $g^r_p(E)$ is the surface retarded Green's
function of semi-infinite lead-p which can be calculated using
transfer matrix method.\cite{transfer} $f_p(E)$ in Eq.(2) is the
Fermi distribution function, it is also a function of the Fermi
energy $E_F$ and temperature $\mathcal{T}$, and can be written as
\begin{equation}
f_{p}(E,E_F^p,\mathcal{T}_p)=\frac{1}{e^{(E-E^p_{F})/k_B\mathcal{T}_p}+1}
\label{f}
\end{equation}
where $E^p_F=E_F+eV_p$ with $e$ the electron charge and $V_{p}$ is
the external bias. In the four terminal system, the thermal
gradient $\Delta \mathcal{T}$ is added between the longitudinal
terminal-1 and terminal-3, and
$\mathcal{T}_1=\mathcal{T}+0.5\Delta \mathcal{T}$,
$\mathcal{T}_3=\mathcal{T}-0.5\Delta \mathcal{T}$, $V_1=V_3=0$.
Due to the Lorentz force, the longitudinal thermal gradient
induces a transverse current $J_{2,4}$ in the closed boundary
condition or a transverse bias $V_{2,4}$ in the open boundary
condition in the terminal-2 and terminal-4. Here we consider the
open boundary ($J_2=J_4=0$) and calculate the balanced bias
$V_{2,4}$. While in the two terminal system, both original thermal
gradient $\Delta \mathcal{T}$ and induced balanced bias are
considered in the longitudinal terminal-1 and terminal-2, and we
have $\mathcal{T}_1=\mathcal{T}+0.5\Delta \mathcal{T}$ and
$\mathcal{T}_2=\mathcal{T}-0.5\Delta \mathcal{T}$. Assuming small
thermal gradient and consequently the small induced external bias,
the Fermi distribution function in Eq.(\ref{f}) can be expanded
linearly around the Fermi energy $E_F$ and the temperature
$\mathcal{T}$,
\begin{eqnarray}
f_p(E,E_F^p,\mathcal{T}_p)&=&f_0+eV_p\left.\frac{\partial
f}{\partial E^p_F}\right|_{V_p=0,\mathcal{T}_p=\mathcal{T}} +\Delta
\mathcal{T}_p\left.\frac{\partial f}{\partial
\mathcal{T}_p}\right|_{V_p=0,\mathcal{T}_p=\mathcal{T}} \nonumber \\
&=&f_0
+f_0(f_0-1)\left[\frac{eV_p}{k_B\mathcal{T}}+(E-E_F)\frac{\Delta
\mathcal{T}_p}{k_B\mathcal{T}^2}\right]
\end{eqnarray}
where $f_0=\left[e^{(E-E_F)/k_B\mathcal{T}}+1\right]^{-1}$ is the
Fermi distribution in the zero bias and zero thermal gradient. Then
for the four terminal system, the current $J_2$ of the terminal-2
can be rewritten as:
\begin{eqnarray}
J_2&=&\frac{2e}{h}\int dE
~f_0(f_0-1)T_{21}(E)\left[(E-E_F)\frac{\Delta
\mathcal{T}}{2k_B\mathcal{T}^2}+\frac{qV_2}{k_B\mathcal{T}}\right]\nonumber
\\ &+&\frac{2e}{h}\int dE~ f_0(f_0-1)T_{23}(E)\left[(E-E_F)\frac{-\Delta
\mathcal{T}}{2k_B\mathcal{T}^2}+\frac{qV_2}{k_B\mathcal{T}}\right]\nonumber
\\ &+&\frac{2e}{h}\int dE~ f_0(f_0-1)T_{24}(E)\left[e\frac{V_2-V_4}{k_B\mathcal{T}}\right]
\end{eqnarray}
Similarly, the expression for the current $J_4$ of the terminal-4
can also be obtained. Using the open boundary condition with
$J_2=J_4=0$ and considering the system symmetry ($T_{21}=T_{43}$,
$T_{23}=T_{41}$ and $T_{24}=T_{42}$), the Nernst coefficient $N_C$
in the four terminal system is:
\begin{eqnarray}
 N_C&=&-\frac{V_2-V_4}{\Delta \mathcal{T}}\nonumber \\
&=&\frac{1}{e\mathcal{T}}\frac{\int
dE~(E-E_F)(T_{21}-T_{23})f_0(f_0-1)}{\int
dE~(T_{21}+T_{23}+2T_{24})f_0(f_0-1)}\label{N}
\end{eqnarray}

In the two terminal system, the current $J_1=-J_2$ is
\begin{eqnarray}
J_1&=&\int dE f_0(f_0-1)T_{21}(E)\left[(E-E_F)\frac{\Delta
\mathcal{T}}{2k_B\mathcal{T}^2}+e\frac{(V_1-V_2)}{k_B\mathcal{T}}\right]\nonumber
\end{eqnarray}
Let $J_1=0$, we have Seebeck coefficient $S_c$
\begin{eqnarray}
 S_C&=&-\frac{V_1-V_2}{\Delta \mathcal{T}}\nonumber \\
&=&\frac{1}{e\mathcal{T}}\frac{\int
dE~(E-E_F)T_{21}(E)f_0(1-f_0)}{\int dE~T_{21}(E)f_0(1-f_0)}\label{S}
\end{eqnarray}

\section{numerical results and discussion}

In the numerical calculations, we set the carbon-carbon distance
$a=0.142nm$ and the hopping energy $t=2.75eV$ as in a real
graphene sample.\cite{ref3,ref4} Throughout this paper the energy
is measured in the unit of $t$. The magnetic field $B$ is
expressed in terms of magnetic flux $BS_0$ in the unit of
$\phi_0/\pi$ where $S_0=\frac{3}{2}\sqrt{3}a^2$ is the area of a
honeycomb unit cell and $\phi_0=\hbar/e$ is the flux quanta. If we
set $BS_0=0.001\phi_0/\pi$, the real magnetic field is around
$4T$. The width of the graphene ribbon is described by an integer
$N$, and the corresponding real width is $3Na$ for zigzag edge
nanoribbon and $\sqrt{3}N a$ for the armchair edge nanoribbon. In
the schematic setup-1I and setup-1II in the inset of Fig.1b,
$N=2$. In the presence of the strong perpendicular magnetic field,
since transport properties are independent of the chirality, we
choose the setup-1I shown in the left inset of Fig.1(b) to study
the Nernst effect and the setup-1II shown in right inset of
Fig.1(b) to study the Seebeck effect. On the other hand, when the
magnetic field is zero, the Seebeck effect strongly depends on the
edge chirality, so we will study both zigzag and armchair edge
nanoribbons, respectively.

\begin{figure}
\includegraphics[bb=10mm 9mm 194mm 222mm,
width=8cm,totalheight=8.5cm, clip=]{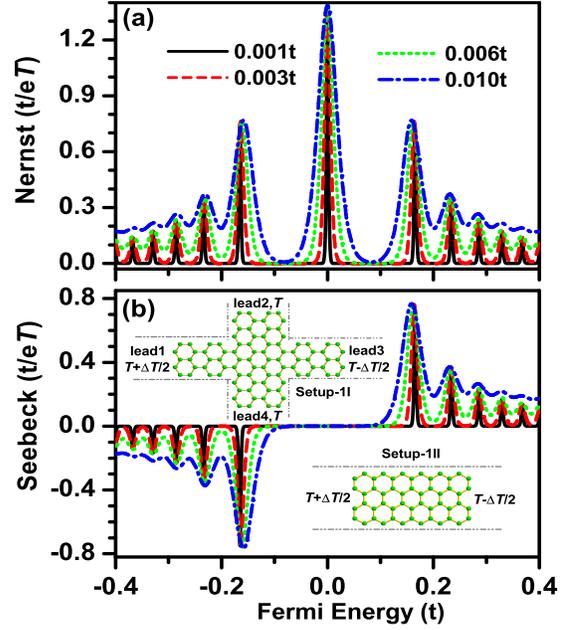} \caption{ (Color
online) Nernst coefficient $N_C$ (a) in the four terminal system and
Seebeck coefficient $S_C$ (b) in two terminal system vs Fermi Energy
$E_F$ with the strong magnetic field $BS_0=0.008\phi_0/\pi$ and
ribbon width $N=80$. Different curves are for different temperatures
$k_B\mathcal{T}$. The four terminal system and the two terminal
system are shown in left and right inset in panel (b),
respectively.} \label{Fig1}
\end{figure}

\subsection{the strong perpendicular magnetic field case}

Firstly, we study the system with strong perpendicular magnetic
field. Fig.1 shows the Nernst coefficient $N_C$ and Seebeck
coefficient $S_C$ versus Fermi energy $E_F$ for different
temperatures $\mathcal{T}=0.001t$, $0.003t$, $0.006t$ and $0.01t$.
Considering the ambipolar nature of the graphene and the
electron-hole symmetry, the Nernst coefficient $N_C$ is an even
function of $E_F$ [$N_C(E_F)=N_C(-E_F)$], because both the energy
$E-E_F$ and the direction of the particle movement (or
$T_{21}-T_{23}$) reverse their signs under the electron-hole
transformation. From Fig.1(a), we see that the Nernst coefficient
$N_C$ show peaks when $E_F$ passes the LLs $E_n=sign(n)\sqrt{2e\hbar
v_F^2|n|B}$ and show valleys between adjacent LLs. With the increase
of the temperature, the peak heights roughly remain unchanged, but
the valleys rise. For convenience, the peaks are numbered and the
peak at $E_F=0$ is denoted as the zeroth peak. In the low
temperature limits, for the $n$-th peak with $n\not=0$, the height
is $[\ln2/|n|]$, and the zeroth peak height is $[2\ln2]$. In
Fig.2(a) we plot inverse of the peak heights versus the peak number
$n$ (see the crossed circle symbols) at the low temperature
$\mathcal{T}=0.001t$. It satisfies the relation $[|n|/\ln2]$. For
comparison, the inverse of the peak's height for the conventional
metal is also plotted (see dotted pentagram symbol), which is
$[(n+1/2)/\ln2]$.

In Fig.1(b), we plot the Seebeck coefficient $S_C$ versus $E_F$ at
different temperatures $\mathcal{T}$. Similarly, the Seebeck
coefficient $S_C$ display peaks when $E_F$ passes the LLs and show
valleys between adjacent LLs. However, $S_C$ shows two essential
differences from the Nernst effect: First, $S_C$ is an odd function
of $E_F$, which means that contributions to $S_C$ from electrons and
holes differ by a sign due to the electron-hole symmetry. So the
Seebeck coefficient $S_C$ is negative for $E_F<0$. Second, when
$E_F$ is on the zero-th LL, $S_C$ is zero instead of the highest
peak in the curve of $N_C$-$E_F$. This is because the zero-th LL
with the fourfold degeneracy is shared equally by electrons and
holes and the electrons and holes give the opposite contributions to
$S_C$. The inverse of the peak height of Seebeck coefficient at the
low temperature ($k_B \mathcal{T}=0.001t$) is plotted in Fig.2(b).
It is found that in graphene, the pseudospin related Berry phase
\cite{ref01} introduces an additional phase shift in the
magneto-oscillation of TEP. As a result of this phase shift, the
inverse of peak height is $\propto n$ (see the crossed circle symbol
in Fig.2(b)). While in the conventional metal or semiconductor with
massive carriers, there is no pseudospin related berry phase, the
inverse of peak height is $\propto n+\frac{1}{2}$ (see dotted
pentacle symbol in Fig.2(b)).

\begin{figure}
\includegraphics[bb=10mm 10mm 192mm 261mm, width=7cm,totalheight=9.0cm, clip=]{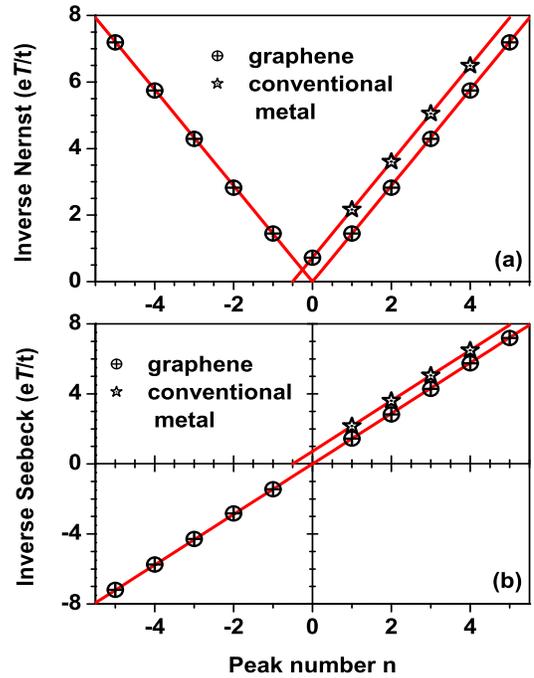}
\caption{ (Color online) The panel (a) and (b) are respectively the
inverse of peak height of Nernst and Seebeck coefficients vs. the
peak number $n$. The crossed circle symbols are for the graphene and
dotted pentagram symbols are for the conventional metal. The
temperature $k_B\mathcal{T}=0.001t$ and other parameters are the
same as Fig.1. In panel (a), the two lines are $|n|/\ln2$ and
$(n+1/2)/\ln2$ and in panel (b) the two lines are $n/\ln2$ and
$(n+1/2)/\ln2$. } \label{Fig2}
\end{figure}

\begin{figure}
\includegraphics[bb=9mm 8mm 178mm 217mm, width=7cm,totalheight=8cm, clip=]{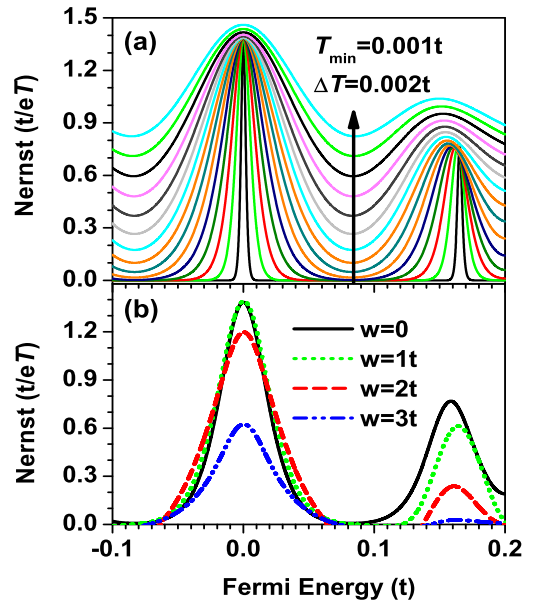}
\caption{ (Color online) Panel (a): the magnifications of the zeroth
and first Nernst peaks in Fig.1(a). Along the arrow direction,
temperature $k_B\mathcal{T}$ increases from $0.001t$ to $0.029t$
with increment of $0.002t$. Panel (b): the disorder effect of panel
(a) at a fixed temperature $k_B\mathcal{T}=0.01t$.} \label{Fig3}
\end{figure}

Next, we study the temperature effect. Since TEP (Nernst effect or
Seebeck effect) represents the entropy transported per unit charge,
both Nernst coefficient and Seebeck coefficient increase with the
increasing temperature which are exhibited in Fig.1(a) and (b). To
take a closer look in Fig.3, we plot the zeroth and first peak for
the temperature range $\in[0.001t,0.029t]$ in the step of $0.002t$.
The temperature effect of $S_C$ is similar to that of $N_C$, so we
only show the Nernst coefficient $S_C$ in Fig.3(a). At low
temperatures, with the increase of the temperature $k_B
\mathcal{T}$, the peak height and position do not vary much, but the
peak half-width is broadened proportional to $k_B \mathcal{T}$, so
the valley between the LLs rises. When the temperature
$k_B\mathcal{T}$ exceeds the spacing of nearest LLs, the Nernst and
Seebeck coefficients $N_C$ and $S_C$ are enhanced in the whole range
of energies including both the peak and valley because of the
overlap of the neighboring peaks. In addition, except for the zeroth
peak, the peak positions for all other peaks shift towards the
zeroth peak.

Now we study the disorder effect on the Nernst and Seeback effect.
To consider the effect of disorder, random on-site potentials
$\delta\epsilon_i$ in the center region are added with a uniform
distribution $[-W/2,W/2]$ with disorder strength W. The data is
obtained by averaging over up to 1200 disorder configurations. It is
known that when the magnetic field is absent, the Seebeck effect is
strongly affected by the disorder, and the peaks are suppressed even
in the small disorder. On the other hand, in the presence of the
strong magnetic field, the Seebeck effect and Nernst effect are
robust to the disorder, because of the existence of the quantized
Landau level. The bigger the sample is (or the stronger the magnetic
field is), the more robust the Nernst effect and Seebeck effect.
Similar to Fig.3(a), in Fig.3(b) we plot the zeroth and first peak
at fixed temperature $k_B\mathcal{T}=0.01t$ with different disorder
strengths. Here sample size ($N=40$) is smaller than that in
Fig.1(a) (in which $N=80$). With the smaller sample size, the zeroth
universal values of peak height $2\ln2$ can still remain until
disorder $W$ is larger than $1t$. For the first peak, the universal
values of height $\ln2/|n|$ remains at $W=0.3t$ and washes out at
stronger disorder. It means that the Nernst peak corresponding to
the lower Landauer level can resist stronger disorders. In fact,
this effect of disorder has been studied for the thermal response to
the charge current\cite{theory1} or to the spin current.\cite{Cheng}
So, in the following, we will focus only on the clean system.

\begin{figure}
\includegraphics[bb=9mm 10mm 194mm 263mm, width=8cm,totalheight=11.0cm, clip=]{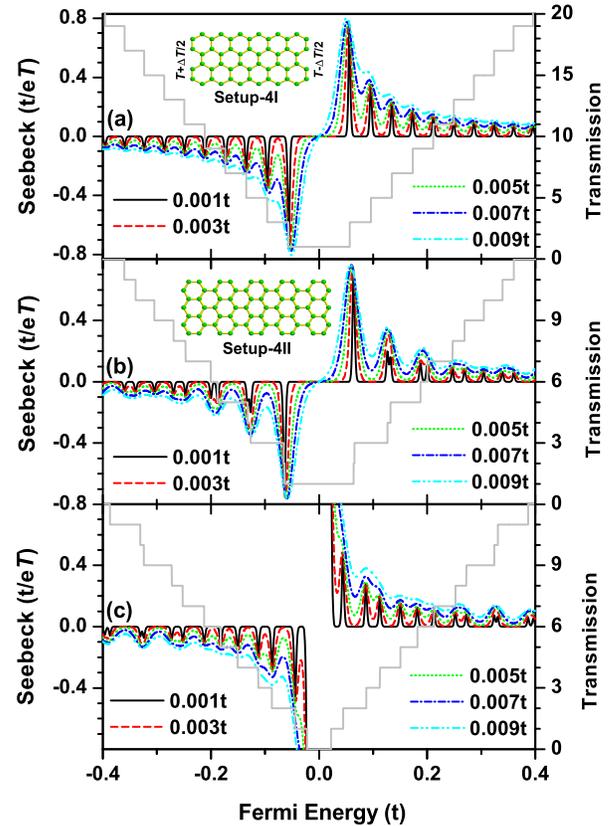}
\caption{ (Color online) Seebeck coefficient $S_C$ vs. Fermi energy
$E_F$ for the different temperatures $k_B\mathcal{T}$ at zero
magnetic field. Panel (a) is for the zigzag ribbon as sketched in
the inset of panel (a), with the width $N=40$. Panel (b) and (c) are
for the armchair ribbon sketched in inset of panel (b) with the
width $N=41$ (b) and $N=40$ (c). The gray solid curves in panels
(a), (b), and (c) are the corresponding transmission coefficients
$T$.} \label{Fig4}
\end{figure}

\begin{figure}
\includegraphics[bb=10mm 11mm 168mm 149mm, width=7cm,totalheight=6.0cm, clip=]{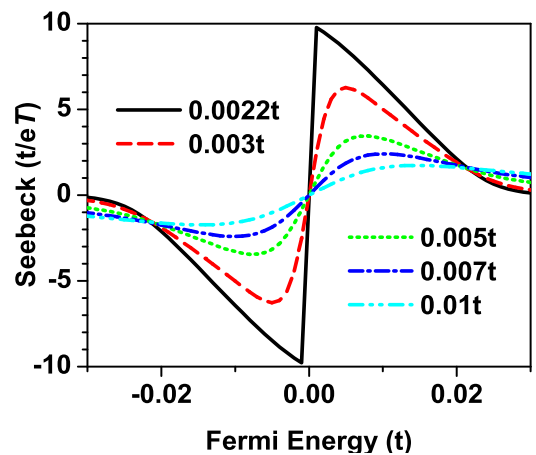}
\caption{ (Color online) The Seebeck coefficient of Fig.4(c) with
the Fermi Energy interval $[-0.03,0.03]$.} \label{Fig5}
\end{figure}

\subsection{the case of zero magnetic field}

In this subsection, we study the TEP at zero magnetic field. Because
there is no Lorentz force to bend the trajectories of the thermally
diffusing carriers, the Nernst effect is absent and $N_C=0$. At
$B=0$, the Seebeck coefficient $S_C$ is strongly dependent on the
chirality of graphene ribbon. In addition, for the armchair edge
ribbon, it is metallic when $N=3M-1$ ($M$ is an integer) and
insulator otherwise.\cite{ref5} The Seebeck coefficient $S_C$ has
essential difference for the metallic and insulator armchair
ribbons. In the following we consider three different systems: (1)
zigzag edge ribbon with width $N=40$ (sketched in inset of
Fig.4(a)), (2) metallic armchair edge ribbon with width $N=41$
(sketched in inset of Fig.4(b)), and (3) insulating armchair edge
ribbon with width $N=40$ (sketched in inset of Fig.4(b)). Fig.4(a),
(b), and (c) show the Seebeck coefficient $S_C$ versus $E_F$ for the
above three systems, respectively. For the convenience of
discussion, we also plot corresponding transmission coefficient
$T=T_{21}=T_{12}$ versus $E_F$ in each panel. We can see that $S_C$
is an odd function of $E_F$ and $S_C$ increases when the temperature
increases. In addition, $S_C$ peaks when Fermi energy crosses the
discrete transverse channels where quantized transmission
coefficient jumps from one step to another. These properties are
similar for the above three cases.

But there are also many essential different behaviors. (1). For the
zigzag edge ribbon, the transverse channels are equidistance with
the energy interval $\Delta=|t|\pi/(2N)$ in the conduction band or
the valence band (except that the interval from the first
transmission channel in the conduction band to the first
transmission channel in the valence band is $3\Delta$). So peaks of
$S_C$ are uniformly distributed over energies and the peak height of
$S_C$ satisfies $[\ln2/2n]$ where $n$ is the peak number (see
Fig.4(a)). (2). In metallic armchair edge ribbon, however, the
transverse channel and consequently the peaks of $S_C$ are
irregularly distributed. The peak height of $S_C$ is closely related
to the transmission coefficient $T=T_{21}=T_{12}$ and it can be
expressed as $2\Delta T ln2/(2T+\Delta T)$ at low temperatures,
where $\Delta T$ is the change of $T$ when $E_F$ scans over the
certain transverse channel. With increasing of the temperature, some
of peaks that are very close to each other merge together so that
both peak height and position are irregular (see Fig.4(b)). (3).
Finally, for the insulating armchair edge ribbon, except for the
irregularly distributed peaks for $|E_F|>\Delta$, the Seebeck
coefficient $S_C$ is very large for $E_F$ near the Dirac point (0)
at low temperatures. Fig.5 magnifies the curves of $S_C$-$E_F$ near
the Dirac point. At low temperatures, $S_C$ can be very large when
$E_F$ approaches the Dirac point. For example $S_C$ can reach about
$10$ at $\mathcal{T}=0.0022t$. At the Dirac point the sign of $S_C$
changes abruptly. This is because near the Dirac point the
transmission coefficient $T_{12}$ is zero and the carriers can't be
transmitted. In order to balance the thermal forces acting on the
charge carriers, we have to add a very large bias leading to a very
large Seebeck coefficient near the Dirac point at low temperatures.
When temperature increase such that $k_B\mathcal{T}$ is greater than
the gap of the insulating armchair edge ribbon $S_C$ decreases
gradually. We emphasize that if the armchair edge ribbon is narrow
enough (such as $W\approx 10nm$ as in our calculation), $S_C\approx
10$ at the temperature $\mathcal{T}=0.0022t/k_B\approx 60K$. This
very large $S_C$ can be observed in the present technology.

\begin{figure}
\includegraphics[bb=9mm 8mm 205mm 176mm, width=8.5cm,totalheight=8.0cm, clip=]{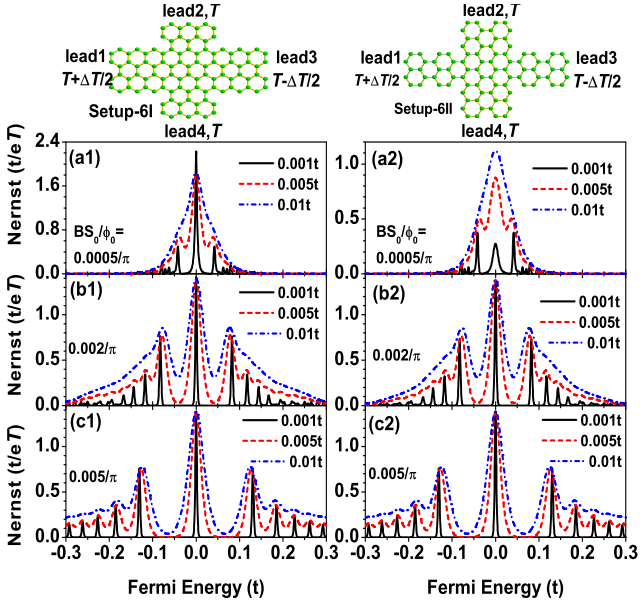}
\caption{ (Color online) Panel (a)-(c) plot the Nernst coefficient
$N_C$ vs Fermi Energy $E_F$ at different temperatures
$k_B\mathcal{T}=0.001t$, $0.005t$ and $0.01t$ in the magnetic field
$BS_0=0.0005\phi_0/\pi,0.002\phi_0/\pi$ and $0.005\phi_0/\pi$,
respectively. In left panels the thermal gradient is added along the
zigzag edge ribbon as shown in the left top sketch. While in the
right panels, the thermal gradient is added along the armchair edge
ribbon as shown in the right top sketch. The ribbon width $N=80$.}
\label{Fig6}
\end{figure}

\subsection{the crossover from zero magnetic field to high magnetic field}

In this subsection, we study the Nernst and Seebeck effect when the
magnetic field varies from zero to finite values (strong magnetic
field). At zero magnetic field, the Nernst coefficient $N_C$ is zero
and the Seebeck effect $S_C$ is dependent on the chirality of
graphene ribbon. At high magnetic fields, however, both $N_C$ and
$S_C$ are independent of the ribbon chirality. What happens with the
magnetic field in the intermediate range?

First, we study the Nernst effect, in which two different setups
(the setup-6I and setup-6II) sketched in the top of Fig.6 are
considered. In Fig.6 we plot the Nernst coefficient $N_C$ versus
$E_F$ at different temperatures and magnetic fields. From Fig.6(a)
to (c), the magnetic field increases from weak to strong enough to
form edge state. At the weak magnetic field (such as
$BS_0/\phi_0=0.0005/\pi$), the Nernst coefficient $N_C$ peaks
sharply near the Dirac point at low temperatures. Because on two
sides of the Dirac point, the carriers are electron-like and
hole-like and they are shifted to the opposite direction under the
weak magnetic field, the Nernst effect is largest at the Dirac
point.

In particular, in the setup-6I, the Nernst coefficient $N_C$ is very
large at the Dirac point, which is much larger than that in
setup-6II and in the case of high magnetic field. Because for the
setup-6I, the longitudinal leads (lead-1 and lead-3) are metallic
with a large transmission coefficient but the transverse leads
(lead-2 and lead-4) are almost insulator near the Dirac point. As a
result, we have to add a much larger bias to balance the thermal
current so that the Nernst coefficient $N_C$ is very large in the
setup-6I at the low magnetic field (see Fig.6(a1) and (b1)). With
increasing of $B$, LLs are formed one by one. The zeroth LL located
at the Dirac point is formed first (at about
$BS_0/\phi_0=0.0015/\pi$, no shown), then is the first LL, the
second and so on. For example, In Fig.6(a), no LL is formed while in
Fig.6(b), the zeroth, first and second LL are formed. As soon as LLs
are formed, the Nernst coefficient $N_C$ will satisfy the relation
that its peak heights are equal to $\ln2/|n|$ (or $2\ln2$ for
$n=0$). From Fig.6(c), we can see that as $BS_0/\phi_0=0.005/\pi$,
electrons (or holes) with Fermi energy $|E_F|\leq 0.3t$ all belong
to robust edge states. In this case, the Nernst coefficient $N_C$
are almost the same for the setup-6I and setup-6II.

For the Seebeck effect, armchair edge ribbon can either be metal or
insulator, we also consider three different systems as in the case
of zero magnetic field. In Fig.7 we plot the Seebeck coefficient
$S_C$ versus $E_F$ at different temperatures and magnetic fields for
three different systems. The first column is for the zigzag edge
ribbon with width $N=80$, the second column is for metallic armchair
edge ribbon with $N=80$, and the third column is for insulating
armchair edge ribbon with $N=81$. From Fig.7(a) to Fig.7(c), the
magnetic field increases gradually. We can see that in the weak
magnetic field, the peaks of $S_C$ are still regularly distributed
for the zigzag ribbon and are irregular for the armchair ribbon due
to the different band structure for the zigzag edge and armchair
edge ribbon. Moreover, for the insulating armchair edge ribbon, the
energy gap near Dirac point is diminished because of the magnetic
field $B$, the very high and sharp $S_C$ at $B=0$ (see Fig.5) is
gradually dropped with the increasing of $B$. But at the weak
magnetic field $BS_0/\phi_0=0.0005/\pi$, the $N_C$ can still reach 3
(see Fig.7(a3)), which is much larger than all peaks of $S_C$ in the
high magnetic field case. Similar to Fig.6, with the increasing of
$B$ further, the LLs is gradually formed from Dirac point to the
high $E_F$, the the properties of $S_C$ for three systems gradually
tend to the same. At the high magnetic field
$BS_0/\phi_0=0.005/\pi$, LLs are completely formed for $|E_F|<0.3$,
then Seebeck coefficient $S_C$ for three different systems are all
the same to that in the Hall region.

\section{conclusion}
In summary, by using the Landauer-B$\ddot{u}$ttiker formula
combining with the non-equilibrium Green's function method, the
Nernst effect in the crossed graphene ribbon and the Seebeck effect
in the single graphene ribbon are investigated. It is found that due
to the electron-hole symmetry, the Nernst coefficient $N_C$ is an
even function while the Seebeck coefficient $S_C$ is an odd function
of the Fermi energy $E_F$. $N_C$ and $S_C$ show peaks when $E_F$
crosses the Landau levels at high magnetic fields or crosses the
transverse sub-bands at the zero magnetic field. In the strong
magnetic field, due to the fact that high degenerated Landau levels
dominate transport processes the Nernst and Seebeck coefficients are
similar for different chirality ribbons. The peak height of $N_C$
and $S_C$, respectively, are $[\ln2/|n|]$ and $[\ln2/n]$ with the
peak number $n$, except for $n=0$. For zeroth peak, it is abnormal.
Its peak height is $[2\ln2]$ for the Nernst effect and it disappears
for the Seebeck effect. While in zero magnetic field, Nernst effect
is absent and the Seebeck effect is strongly dependent on the
chirality of the ribbon. For the zigzag edge ribbon, the peaks of
$S_C$ are equidistance, but they are irregularly distributed for
armchair edge ribbon. Surprisingly, for the insulating armchair edge
ribbon, the Seebeck coefficient $S_C$ can be very large near the
Dirac point due to the energy gap. When the magnetic field increases
from zero to high values, the irregularly or regularly distributed
peaks of $S_C$ in different chiral ribbons gradually tends to be the
same. In addition, the nonzero values of the Nernst coefficient
$N_C$ appear first near the Dirac point and then gradually in the
whole energy region. It is remarkable that for certain crossed
ribbons, the Nernst coefficient $N_C$ at weak magnetic fields can be
much larger than that in the strong magnetic field due to small
transmission coefficient in the transverse terminals.

$${\bf ACKNOWLEDGMENTS}$$

We gratefully acknowledge the financial support by a RGC grant (HKU
704308P) from the Government of HKSAR and NSF-China under Grants
Nos. 10525418, 10734110, and 10821403.

\newpage
\begin{figure}
\includegraphics[bb=9mm 9mm 208mm 136mm, width=12cm,totalheight=8.0cm, clip=]{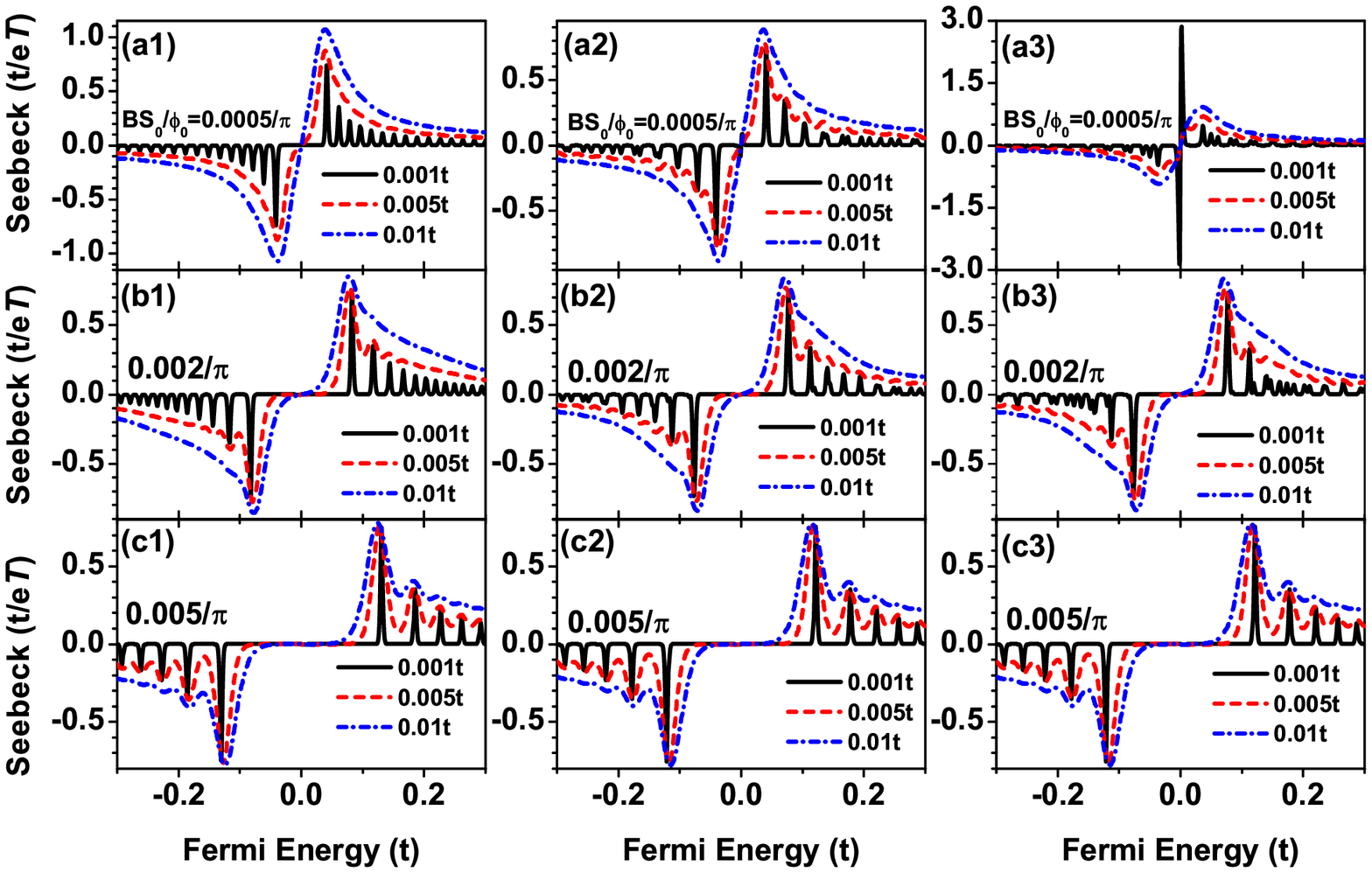}
\caption{ (Color online) Panel (a)-(c) plot the Nernst coefficient
$N_C$ vs Fermi Energy $E_F$ at different temperatures
$k_B\mathcal{T}$ and different magnetic fields $BS_0/\phi_0$. The
other parameters and the chirality of ribbon for the first, second,
and third column panels are the same as Fig.4(a), (b), and (c),
respectively.} \label{Fig7}
\end{figure}

\end{document}